\documentclass[aps,pra,reprint,superscriptaddress]{revtex4-1}
\pdfoutput=1
\usepackage{amsmath,graphicx,hyperref,color}
\usepackage[FIGTOPCAP,raggedright,nooneline,bf,footnotesize]{subfigure}
\usepackage{indentfirst}
\usepackage{braket}
\usepackage{multirow}
\usepackage{hyperref}

\renewcommand{\eqref}[1]{Eq.~(\ref{#1})}

\begin{document}

\title{Interaction of a heralded single photon with nitrogen-vacancy centers in diamond}

\author{Maria Gieysztor}
\author{Marta Misiaszek}
\affiliation{Institute of Physics, Faculty of Physics, Astronomy and Informatics, Nicolaus Copernicus University in Toru\'{n}, Grudziadzka 5, 87-100 Toru\'{n}, Poland} 

\author{Joscelyn van der Veen}
\affiliation{Faculty of Physics, Astronomy and Informatics, Nicolaus Copernicus University, Grudziadzka 5, 87-100 Toru\'{n}, Poland} 
\affiliation{Department of Physics and Astronomy, University of Waterloo, 200 University Ave W, Waterloo, Ontario, N2L 3G1, Canada} 

\author{Wojciech Gawlik}
\affiliation{Institute of Physics, Jagiellonian University, \L{}ojasiewicza 11, 30-348 Krak\'{o}w, Poland} 

\author{Fedor Jelezko}
\affiliation{Institute for Quantum Optics, University of Ulm, Albert-Einstein-Allee 11, D-89081 Ulm, Germany} 

\author{Piotr Kolenderski}
\email{kolenderski@fizyka.umk.pl}
\affiliation{Faculty of Physics, Astronomy and Informatics, Nicolaus Copernicus University, Grudziadzka 5, 87-100 Toru\'{n}, Poland} 

\date{\today}

\begin{abstract}
	
\noindent
A simple, room-temperature, cavity- and vacuum-free interface for photon-matter interaction is implemented. In the experiment a heralded single photon generated by the process of spontaneous parametric down-conversion is absorbed by an ensemble of nitrogen-vacancy color centers. The broad absorption spectrum associated with the phonon sideband solves the mismatch problem of a narrow absorption bandwidth in a typical atomic medium and broadband spectrum of quantum light. The heralded single photon source is tunable in the spectral range $452-575$ nm, which overlaps well with the absorption spectrum of nitrogen-vacancy centers. 

\end{abstract}


\maketitle

\section{Introduction}
Efficient control of light-matter interaction at a single-particle level is a key factor enabling several quantum applications. One of the main goals in quantum technology is to deliver a quantum network platform for secure quantum communication. Atom or trapped ion implementations of quantum memories and quantum repeaters together with photonic information carriers are one of the proposed scenarios \cite{Wang2017, Piro2011, Brito2016}. Moreover, quantum imaging and sensing benefit from single-photons interaction with matter. In particular, this is apparent when discussing quantum illumination, sub-shot-noise imaging, ghost imaging, and absolute detector calibration \cite{Meda2017}. Further, absorption microscopy exhibts higher signal to noise ratio when quantum light is applied \cite{Li2018}. Entangled two-photon absorption is another microscopy technique which has a great potential to outperform its classical counterpart in terms of applicability \cite{Simon2016}. It may also serve as a tool for virtual-state spectroscopy \cite{J.Leon-Montiel2019,Svozilik2018,Dorfman2012}. However, all the above-mentioned applications require an efficient single-photon and single atomic system interaction, which still remains a challenge. The narrow atom absorption spectrum together with spectrally broad single-photon sources demand cavities to enhance the interaction efficiency \cite{Cirac1997, Ritter2012, Specht2011}. 

Here, for the first time, a different solution is proposed and the mismatch problem is solved by using a spectrally broad absorber, such as a color center in diamond, where a nonresonant excitation is possible due to phononic sidebands \cite{Doherty2013}. This eliminates the need for excitation with a narrow bandwidth light. A plethora of such systems, like SiV, GeV, SiC or dyes used in biological imaging, is already known and investigated \cite{Becker2017a, Neu2017, Bosma2018}. A simple, room-temperature, cavity- and vacuum-free interface for photon-matter interaction is reported. The heralded single-photon source (HSPS) is based on the spontaneous parametric down conversion (SPDC) process, where the detection of an infrared photon is used as a herald for the visible one. In the experiment, a heralded single photon is absorbed by a single atom-like system, specifically a nitrogen vacancy center (NV) in diamond. The NV center then emits another photon, whose arrival time is measured by means of a time-resolved single-photon detection technique. As opposed to Ref.~\cite{Fisher2017} we do not use a stimulated absorption-emission process and our method is single atom oriented. 

\section{Experimental setup}

\begin{figure*}[htb!]
	\includegraphics[width=\textwidth]{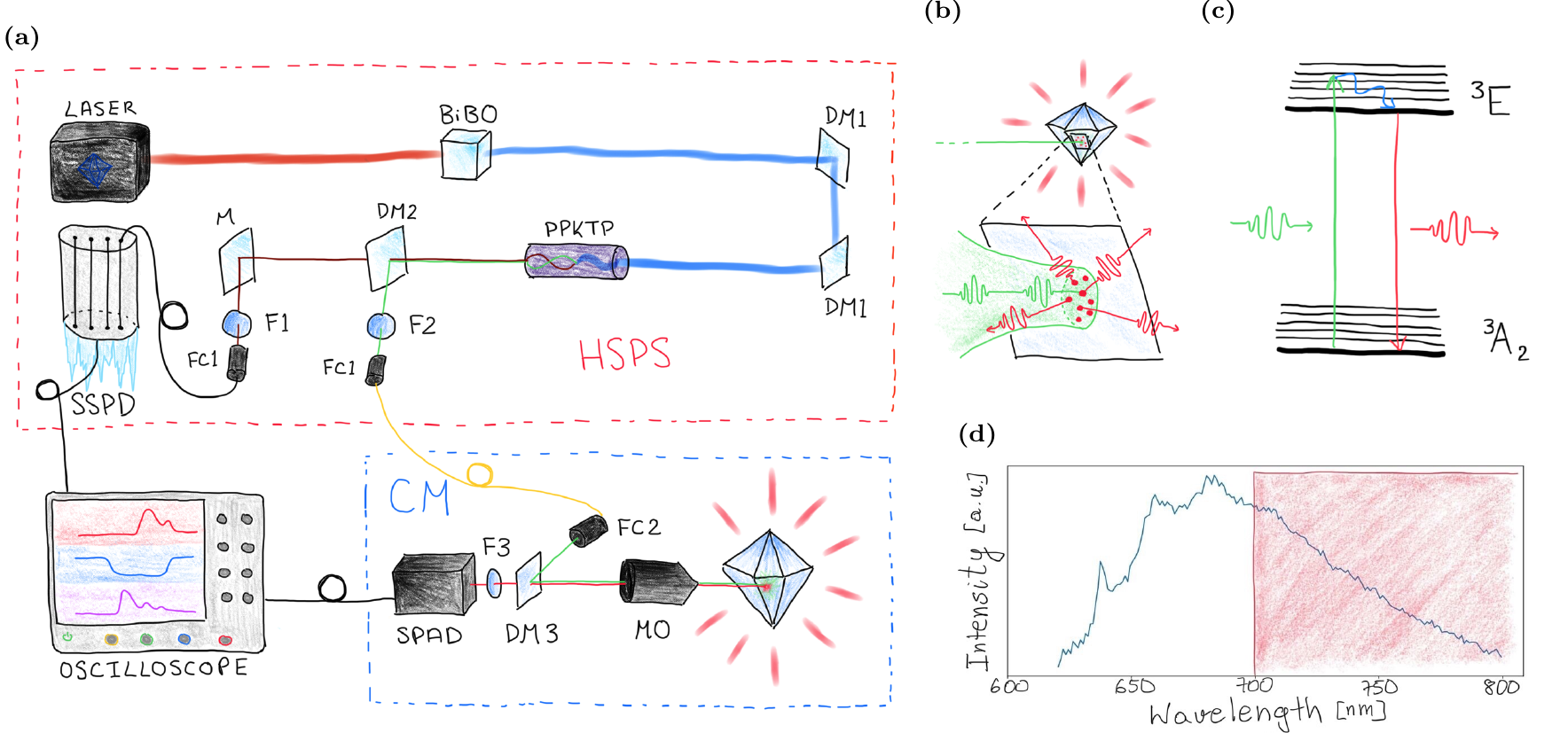}
	\caption{Interaction between photons and NV centers. (a) The experimental setup.  (b) Magnification of the diamond surface: One of the NV centers from the  illuminated ensemble absorbs a heralded green photon and then emits a fluorescence photon. (c) Simplified energy level structure of an NV centre: a green photon excites an electron from the ${}^3A_2$ (ground) state to a ${}^3E$ (excited) state. The electron then decays non-radiatively to a lower energy excited state, transferring some of its energy to the diamond lattice. Finally, the electron decays radiatively back to the ${}^3A_2$ (ground) state via the emission of a lower energy red photon. (d) Blue: NV$^-$ fluorescence spectrum with zero phonon line (ZPL) at $637$ nm and vibronic sideband up to $800$ nm. Red: the transmission range of the spectral filtering setup (DM3 and F3). }
	\label{fig:setup}
\end{figure*}

The experimental setup is depicted in Fig. \ref{fig:setup} (a).  A red laser photon beam incident on the BiBO crystal is frequency doubled. Then a blue pump photon is converted through the SPDC process into a pair of photons -- one in visible and one in infrared spectral range. When the superconducting single photon detector (SSPD) measures the latter, the existence of the former is heralded and the time reference is defined. Subsequently, the heralded visible photon is delivered to the custom-made confocal microscope (CM), where it is reflected off a dichroic mirror (DM) and focused with a microscope objective (MO) onto an ensemble of color centres. The sample used in the experiment is a high-pressure high-temperature (HPHT) diamond with a dense concentration of negatively charged nitrogen-vacancy (NV$^-$) centres, approximately $18$ ppm. Full sample characterization and preparation description, including spectrum and ODMR measurement, is given in Refs.~\cite{Mrzek2015, Rudnicki2013}. Next, the photon excites a color centre, which results in a fluorescence photon emission in the range of $600-800$ nm  \cite{Aslam2013}, see Fig.~\ref{fig:setup} (d).  The resulting photon is collected by the same MO and propagates through the DM and a longpass filter. This allows observation of fluorescence only in the spectral range above $700$ nm, which corresponds to the emission of the negatively charged NV centers mainly \cite{Aslam2013}. Finally, its arrival time is measured by a single-photon avalanche diode (SPAD)  supported by a digital oscilloscope.  The statistics of arrival times collected by the measurement system are used to build a histogram exhibiting the characteristic fluorescence decay shape.

The HSPS used in the experiment is tunable in the range of $452$-$575$ nm \cite{Divochiy2018}. The FWHM of the HSPS in the visible spectrum is around $4$~nm, which corresponds to $3.2$~THz at $575$~nm and $5.9$~THz at $450$~nm. The scenario with a $532$ nm single-photon was chosen but other pumping scenarions were tested as well giving similiar results. Two pumping power settings, 1 and 5 mW, were used for the source, which are referred to as lower and higher, respectively. The high quality of heralded single photon state was verified by measuring the correlation function $g^{(2)}(0)$ \cite{Glauber1963a, URen2005a}, describing the goodness of the single-photon source. In the perfect case it is $0$ for a single photon Fock state and $1$ for attenuated laser described by a coherent state. Here, for lower power pump it took the value of $g^{(2)}_{l}(0)=0.0011(2)$, whereas for higher pump power $g^{(2)}_{h}(0)=0.0111(5)$. The corresponding heralded photon count rates were $4.5$~kcps (lower power) and $40$~kcps (higher power). The HSPS characterization was done for $532$ nm.

\section{Fluorescence decay measurement}

\begin{figure}[htb!]
	\centering
	\includegraphics[width=6.5cm]{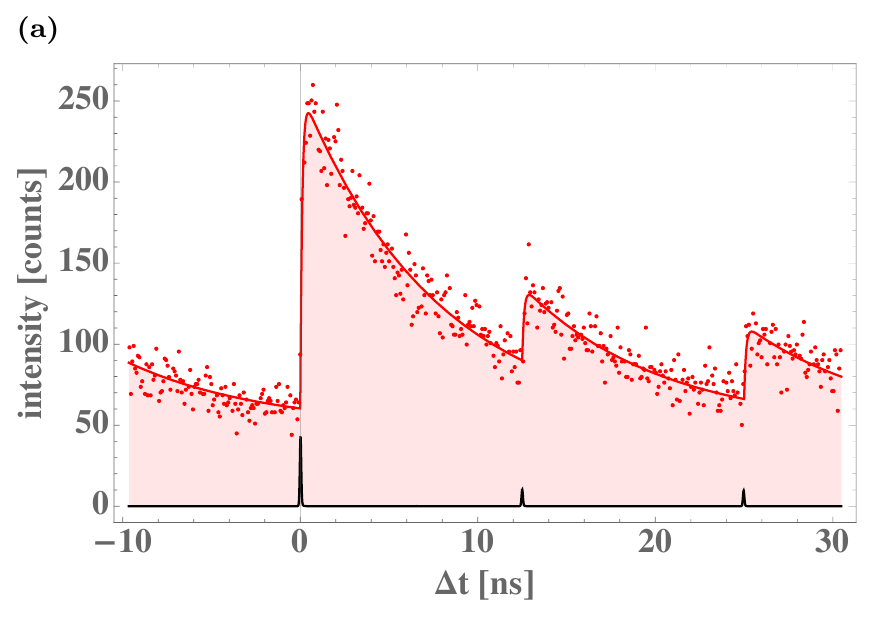}
	\includegraphics[width=6.5cm]{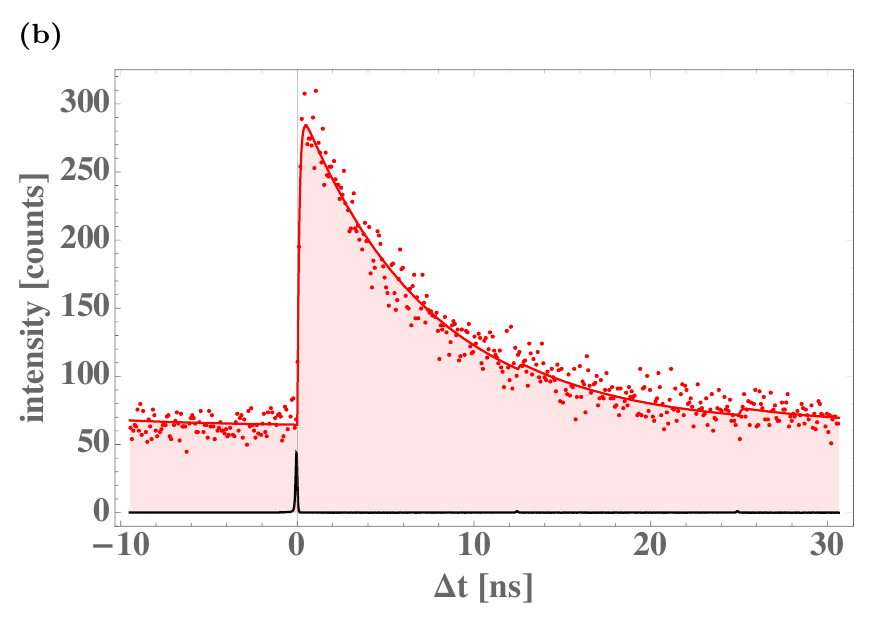}
	\caption{NV$^-$ fluorescence decay measurement pumped with single photons. Single photons with wavelength of $532$ nm are generated in the SPDC process in (a) higher and (b) lower pumping power setting. Red dots show experimental data and red curve represents fitted model. Black curve shows a histogram of visible and infrared (IR) photon coincidences measured from HSPS source. Large and small peaks correspond to proper and accidental coincidences, respectively. The accidental coincidences almost disappear in the lower power setting. More detailed information can be found in Supplementary Information.}
	\label{fig:decays_fit}
\end{figure}

In Fig.~\ref{fig:decays_fit} the obtained NV fluorescence decay pumped with heralded single photons is shown for the two pumping power settings. The probability of the fluorescence emission in time, $p(t)$, can be described by a simple exponential model including a non-radiative decay path. The model assumes a fast non-radiative decay followed by a radiative decay, with characteristic times $\tau_N$ and $\tau_R$, respectively. The explicit formula can be obtained by taking a convolution of the two single exponential decays

\begin{equation}
	p(t) = \theta(t) \cdot \frac{1}{\tau_N - \tau_R} 
	\left(
	e^{-\frac{t}{\tau_N}} - e^{-\frac{t}{\tau_R}}
	\right), 
	\label{eq: model_single_conv}
\end{equation}
where $\theta(t)$ is the step-function (equal to $0$ for $t<0$ and to $1$ for $t\geq0$). The SPDC source is pumped with a laser emitting pulses of $140$~fs time duration every $t_0 = 12.5$~ns. Therefore, a photon pair can be generated every multiple of the repetition period $t_0$. Note, that due to the probabilistic nature of the SPDC process not every pulse generates a pair. On top of that due to the losses in the experimental setup not all of the observed coincidences registered in the detection system originate from the same pump pulse. Sometimes the pulse that produces the visible photon exciting the NV center and causing the fluorescence is different than the pulse from which the detected heralding photon is created. Such coincidences are referred to as accidental, while the coincidences caused by a single pump pulse are called proper. The ratio of accidental to proper coincidences, $r$, can be also used as a parameter of the fluorescence model given by the formula below
\begin{equation}
	P(t)  =   a \left(  p(t) + r \sum_{n=1}^{\infty} \left[ p(t + n t_0) + p(t - n  t_0) \right] \right)  + b ,
	\label{eq: model_single_conv_final}
\end{equation}
where $a$ stands for the coefficient scaling the normalized probability given in \eqref{eq: model_single_conv} to the actually detected counts and $b$ reflects the background originating from the dark counts. The  details of the derivation of \eqref{eq: model_single_conv} and \eqref{eq: model_single_conv_final} can be found in Supplement.

The radiative, $\tau_R$, and non-radiative, $\tau_N$, decay times  along with the ratio $r$ can be estimated by fitting  the model to the experimental data. Their values for higher (lower) pumping setting were determined to be  $7.68(23)$ ns and $107(14)$ ps ($7.17(14)$ ns and $112(13)$ ps), respectively. The obtained radiative decay times are shorter with respect to what is reported in the literature \cite{Batalov2008}. Two possible explanations for this discrepancy are proposed. Firstly, the sample is illuminated with single photons resulting in the NV center not being spin-polarized, as opposed to the case of the excitation with a high power laser \cite{Batalov2008,Robledo2011}.  Secondly, the experiment is performed on a very dense sample, where the shorter decay time can be attributed to the F\"orster resonance energy transfer (FRET). It makes the fluorescence emission faster when the emitters are close to each other \cite{Work1969}. Next, the fitted ratio, $r$, takes the value of $0.233(9)$ ($0.0225(81)$) in the higher (lower) pumping power setting. The difference reflects the fact that for the higher pump power setting of the HSPS the ratio of accidental to proper coincidences is higher. It can be clearly seen when comparing black histograms in Fig.~\ref{fig:decays_fit}.

\begin{figure}[htb!]
	\centering
	\includegraphics[width=7cm]{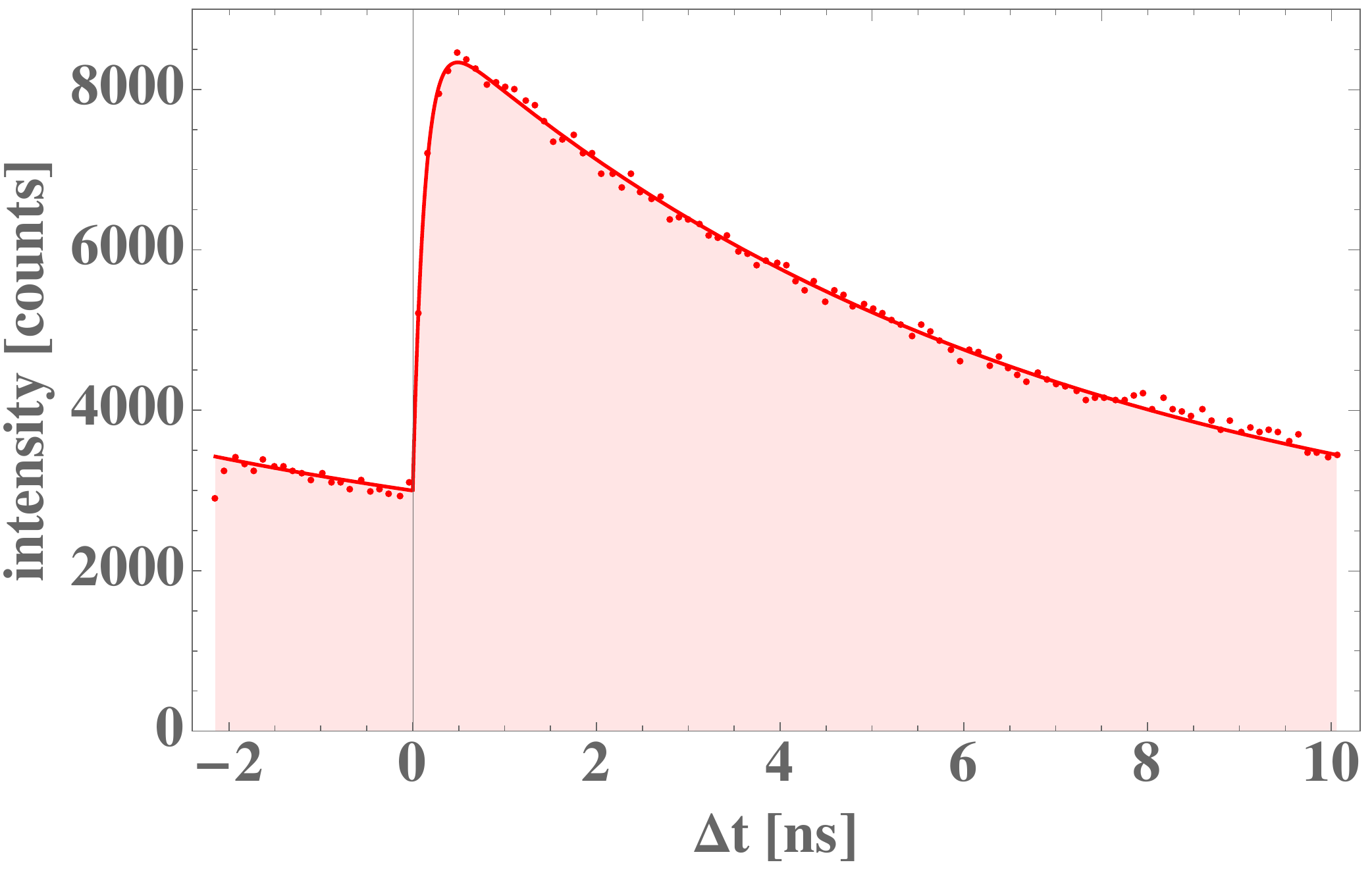}
	\caption{NV$^-$ fluorescence decay measurement pumped with puled attenuated laser. The pumping wavelength was $532$ nm, pulse duration approximately $200$ fs and repetition rate $80$ MHz.}
	\label{fig:coherentDECAY}
\end{figure}

As a reference measurement, an experiment was performed with the heralded single photons replaced with an highly attenuated, pulsed laser. The histogram of the measured photon detection times with respect to the optical laser pules is given in Fig. \ref{fig:coherentDECAY}. A fluorescence lifetime model, which takes into account repetitive excitation, was used for fitting. The obtained radiative and non-radiative dacay times equalled $\tau_R = 6.6 (2) $ ns and $\tau_N = 127 (8) $ ns, respectively. 

\section{Discussion}

The detector monitoring the fluorescence has an active area of $50$ $\mu$m diameter, photon detection efficiency of the order of $15-28$ \% for the spectral range of interest, which is $700-800$ nm, and exhibits $65(8)$ dark counts per second. When taking into account the number of fluorescence photons, the signal to noise ratio (SNR) for the higher (lower) power setting was $0.65(46)$ ($0.154(16)$). However, the heralding scheme applied in the experiment, taking into account only the heralded fluorescence photons and dark counts, improves SNR because it benefits from the ability to reject a substantial part of the dark counts. It stems from the fact that the detection window for the fluorescence monitoring detector is being opened only when a heralding photon was already detected.  Hence, the effective detection probability of a fluorescence photon is higher than of a dark count within a given detection window. For experimental conditions in the higher (lower) power setting this results in $2.96(27)$\% ($1.42(16)$\%) of the real fluorescence and $0.446(12)$\% ($0.172(22)$\%) of the dark counts being heralded. This leads to the increase of SNR for the higher (lower) power setting to $4.3(8)$ ($1.3(6)$). In principle, the SNR can be improved even further by replacing the fluorescence monitoring by one with active area of $30$ $\mu$m diameter giving and reduced dark count rate $1$ per second \cite{Sanzaro2018}. The expected SNR would increase by a factor of $65$. Details on SNR calculation can be found in Supplement.

Now the problem of converting a heralded photon generated by HSPS to the fluorescence photon emitted by an NV center is adressed. This quantity can be estimated based on the observed count rates to be $1.46(32) \cdot 10^{-4}$ ($2.1(8) \cdot 10^{-4}$) for higher (lower) power setting. This is the efficiency of the sample, meaning the experimental setup imperfections are excluded. The setup sets additional limits resulting in the observed conversion efficiency, $\eta_{conv.}$, of $7.0(7)\cdot 10^{-6}$ ($1.01(25)\cdot 10^{-5}$) for higher (lower) pumping setting. This is due to experimental factors including: efficiency of extracting fluorescence photons from the diamond, losses in optical elements, fiber coupling efficiencies and quantum efficiencies of the detectors.

The low conversion efficiency results in long time of data acquisition. The typical time of the reported experiments was of the order of 24 hours. The heralded fluorescence photon count rate (signal), $n_{sh}$, depends on the SPDC generation rate, $N_{SPDC}$, visible photon fiber coupling, $\eta_{VIS}$, infrared photon fiber coupling, $\eta_{IR}$, infrared photon detection efficiency, $\eta_{IR det.}$, and the conversion efficiency, $\eta_{conv.}$. The exact formula is a product of those factors, $n_{sh} = N_{SPDC} \cdot \eta_{VIS} \cdot \eta_{IR} \cdot \eta_{IR det.} \cdot \eta_{conv.}$. Hence it is clear, that improvement of the SPDC photon pair coupling efficiency, increasing the SPDC source pomp power, and detectors with higher quantum efficiency can speed up the measurement. It is estimated, that when switching from pulsed to continuous-wave laser pumping of the SPDC source, the power can be safely increased up to single Watts with preserved single photon characteristics of the source. This would result in 3 orders of magnitude faster measurement when leaving all other settings unchanged. It is also worth mentioning that while the experiment was performed with the aid of an oil-immersion MO, it is possible to conduct it with a dry air MO. Such configuration would result in longer data aquisition time but would enable cryogenic experiments.

\section{Summary}

In conclusion, a simple, room-temperature, cavity- and  vacuum-free  single-photon single atom-like system interaction was demonstrated. The experiment was performed on a high NV$^-$-concentration diamond sample with a tunable, SPDC-based heralded single-photon source. Interaction efficiency, usually limited by the very small spectrum overlap between the broadband quantum light and narrow atomic transitions, was significantly enhanced thanks to phonon-broadened absorption in nitrogen-vacancy centers in diamond.

The result can also be considered as a useful technique paving the way for development of new applications like  quantum microscopy \cite{Li2018,Simon2016} or virtual-state spectroscopy \cite{J.Leon-Montiel2019,Svozilik2018,Dorfman2012}.  In particular, the ability of interaction of NVs with statistics-controlled quantum light enables two-photon processes, such as ionization of negatively-charged NV centers, to be eliminated. Moreover, it can be extended to the scenario of a low NV$^-$ concentration sample, which would enable addressing of a single color centers. It would facilitate testing of the fundamental properties of quantum light interacting with atomic systems. As a further step, several applications, such as quantum microscopy, can be enhanced when non-classical properties of quantum light are exploited.

See Supplement for supporting content.

\section*{Acknowledgments}
MG, MM, PK acknowledge financial support by the Foundation for Polish Science (FNP) (project First Team co-financed by the European Union under the European Regional Development Fund). Equipment was partially financed by  Ministry of Science and higher Education, Poland (MNiSW) (grant no.~6576/IA/SP/2016) and National Science Centre, Poland (NCN) (Sonata 12 grant no.~2016/23/D/ST2/02064). This research has been partially financed from the funds of the Polish Ministry of Science and Higher Education for statutory R\&D activities supporting the development of young scientists and PhD students (internal grant no. 1036-F/2018, 1040-F/2018). WG acknowledges financial support by the Foundation for Polish Science (FNP) (project Team grant no. Net POIR.04.04.00-00-1644/18) and by the National Science Centre, Poland (NCN) (Opus 11 grant no.~2016/21/B/ST7/01430). The authors thank Adam Wojciechowski, Mariusz Mr\'{o}zek, and Andrzej Kruk for fruitful discussions.

\end{document}


\title{Interaction of a heralded single photon with nitrogen-vacancy centers in diamond\\ Supplementary Information}

\author{Maria Gieysztor}
\author{Marta Misiaszek}
\affiliation{Institute of Physics, Faculty of Physics, Astronomy and Informatics, Nicolaus Copernicus University in Toru\'{n}, Grudziadzka 5, 87-100 Toru\'{n}, Poland} 

\author{Joscelyn van der Veen}
\affiliation{Faculty of Physics, Astronomy and Informatics, Nicolaus Copernicus University, Grudziadzka 5, 87-100 Toru\'{n}, Poland} 
\affiliation{Department of Physics and Astronomy, University of Waterloo, 200 University Ave W, Waterloo, Ontario, N2L 3G1, Canada} 

\author{Wojciech Gawlik}
\affiliation{Institute of Physics, Jagiellonian University, Lojasiewicza 11, 30-348 Krak\'{o}w, Poland} 

\author{Fedor Jelezko}
\affiliation{Institute for Quantum Optics, University of Ulm, Albert-Einstein-Allee 11, D-89081 Ulm, Germany} 

\author{Piotr Kolenderski}
\email{kolenderski@fizyka.umk.pl}
\affiliation{Faculty of Physics, Astronomy and Informatics, Nicolaus Copernicus University, Grudziadzka 5, 87-100 Toru\'{n}, Poland} 

\date{\today}

\pacs{}

\maketitle


\section{Fluorescence decay model} 
\label{appendix:deriv}

When the NV centre absorbs a photon, it non-radiatively decays to a lower energy excited state, before decaying further to the ground state and emitting a photon. Since the two processes occur consecutively, the fluorescence emission probability over time, $p(t)$, can be calculated as a convolution of the non-radiative, $p_N(t)$, and radiative, $p_R(t)$: decay probability distributions
\begin{equation}
p(t) = \int_{0}^{t}p_N(t')p_R(t-t')dt' \\
= \frac{1}{\tau_N - \tau_R} \left(e^{-\frac{t}{\tau_N}} - e^{-\frac{t}{\tau_R}}\right),
\end{equation}
where the respective function are defined as:
\begin{equation}
p_N(t) = \frac{1}{\tau_N}e^{-t/\tau_N}, \\
p_R(t) = \frac{1}{\tau_R}e^{-t/\tau_R}.
\end{equation}
Only the probability of the decay for $t > 0$ is considered since the NV centre must be excited before it can decay. Thus, the probability distribution $p(t)$ is redefined to

\begin{equation}
p(t) = \theta(t) \cdot \frac{1}{\tau_N - \tau_R}\left(e^{-\frac{t}{\tau_N}} - e^{-\frac{t}{\tau_R}}\right),
\label{eq: model_single_conv:redefined}
\end{equation}
where $\theta(t)$ is the step-function equal to zero (one) for the negative (positive) values of t such that the probability of decay is $0$ for $t < 0$.

\begin{figure*}
	\centering
	\includegraphics[width=0.85\textwidth]{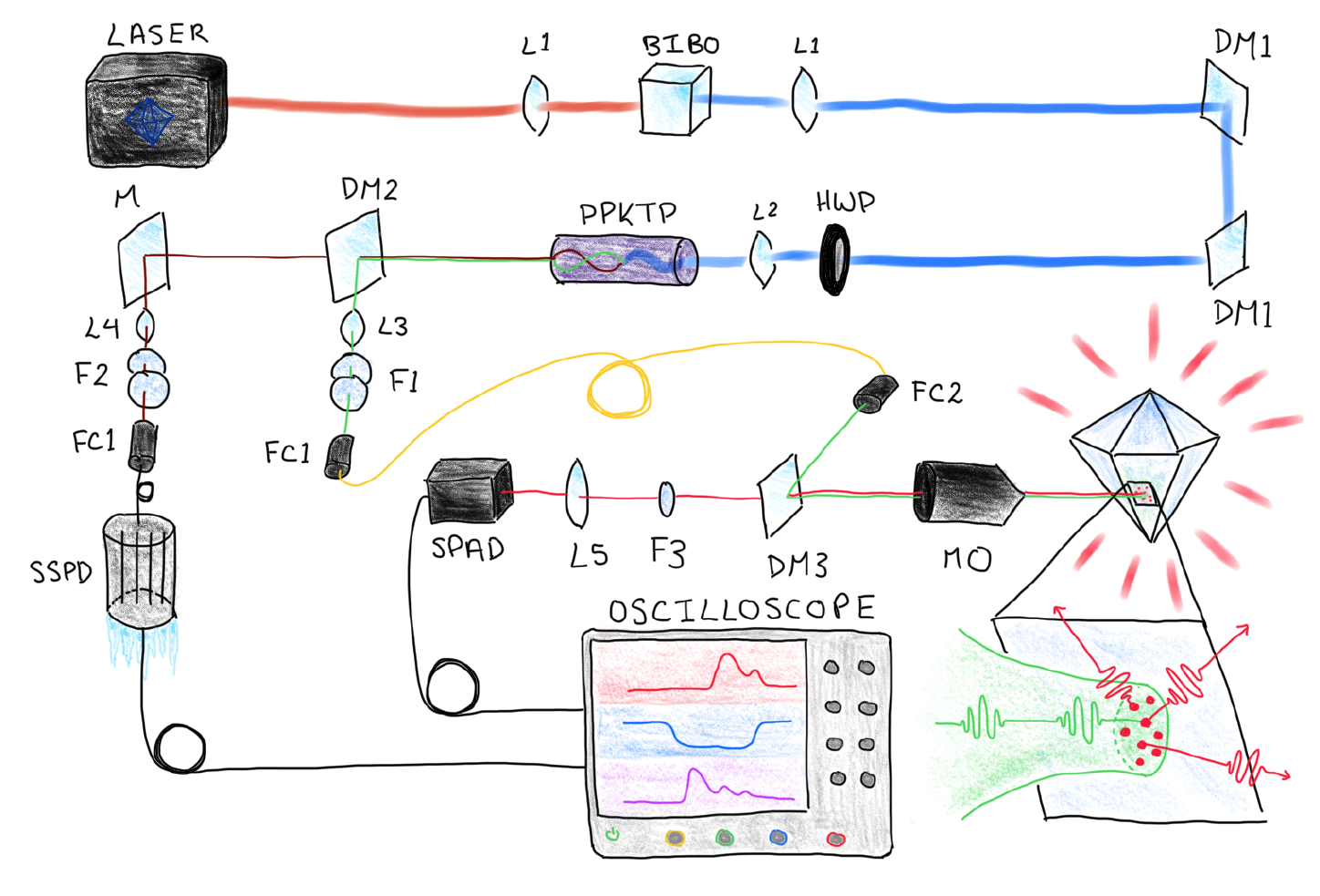}
	\caption{Detailed experimental setup. Infrared light from a Ti:Sapphire pulsed laser is focused by lens L1 (plano-convex, f $= 7.5$ cm) onto a nonlinear BiBO crystal. The resulting blue photons are collimated by another lens L1. Two dichroic mirrors, DM1 (Chroma AHF T425LPXR), separate unconverted laser light. The halfwave plate HWP rotates the laser beam polarization before it is focused by lens L2 (plano-convex, f $= 10.0$ cm) into the type-II PPKTP nonlinear crystal ($10\times4\times2 \mu m$). Dichroic mirror DM2 (Chroma AHF $76-875-$LP) separates the visible and infrared photons. Lens L3 (plano-convex, f $= 15.0$ cm) collimates the visible photons and then filters F1 (ET$500$ and FF$550/88$) remove unconverted laser light before photons are coupled into a single mode fibre (SMF780) by fibre collimation package with focal length either f $= 1.51$ cm or f $= 1.1$ cm (FC1). Infrared photons are transmitted through DM2, collimated by lens L4 (plano-convex, f $=10.0$ cm), filtered by filter F2 ($1319$ LP), and coupled into a single mode fibre (F1-$2000$-FC-1) using mirror M and fibre collimator FC1 (f $= 1.51$ cm). Superconducting nanowire single photon detector, SSPD, detects the IR photons. Visible photons (532/565/575 nm) travel through fibres and are collimated by collimation package FC2 (F$671$FC$-405$) to reflect off dichroic mirror DM3 (Semrock FF$573$-DIO1). Microscope objective MO focuses green photons on diamond sample and collects red fluorescence. Red photons pass through DM3 and remaining green photons are removed by filter F3 (Thorlabs FELH550/FELH700/BP$650-40$). For tests with 532 nm, 565 nm, and 575 nm green photons, fluorescence photons with wavelengths below 550 nm or 700 nm, below 700 nm, and between 610 nm and  690 nm were filtered respectively. Lens L5 (plano-convex, f $= 3.0$ cm) focuses photons onto avalanche photodiode SPAD. Oscilloscope detects signals from SSPD and MPD.}
	\label{fig:fullsetup}
\end{figure*}

\begin{figure}[t!]
	\centering
	\includegraphics[width=0.99\columnwidth]{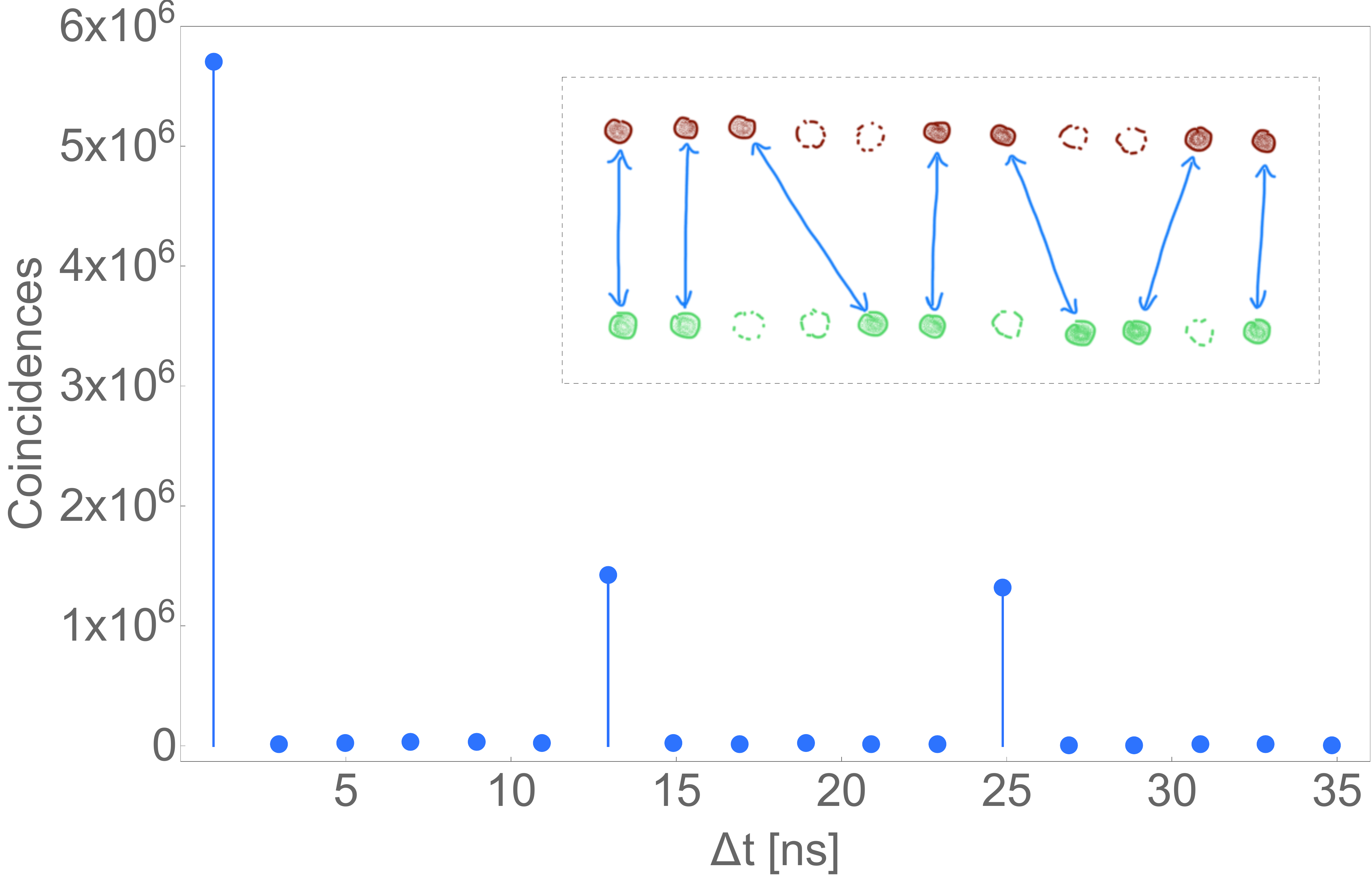}
	\caption{Accidental coincidences. Time delay between visible and infrared SPDC photons for higher and lower power settings, respectively. The accidental coincidences almost disappear in the lower power setting. The inset visualizes the concept: when one of the photons in the SPDC photon pair is lost (dashed, empty circles), this can result in coincidences between photons in neighbouring pairs (solid, colored circles), accidental coincidences.}
	\label{fig:acc_coinc}
\end{figure}

The detailed version of the experimental setup scheme for the heralded measurement illustrated in Fig. 1 (a) in the main text is given in \figref{fig:fullsetup}. In the SPDC process, a pair of photons is produced in a nonlinear crystal. Since the photons are created together and the chromatic dispersion in the setup in negligible, the time delay between their detection events is constant and depends on the difference in path lengths of the photons and detection signal. In a histogram of time delays, Fig. \ref{fig:acc_coinc}, this is visible as a peak in coincidences, further called proper coincidence peak. However, due to setup imperfections, one or both of the photons from a given pair can be lost. When this occurs, the coincidence may instead happen between photons from different pairs, as shown in Fig. \ref{fig:acc_coinc}~inset. This results in a time delay that is a multiple of the rate of generation of the photon pairs, which is the time between laser pulses $t_0$. These events, called accidental coincidences, can be seen as smaller peaks in the number of coincidences in Fig. \ref{fig:acc_coinc}.

The presence of accidental coincidences has a direct influence on the heralded measurement and hence further modifiactions of the analytical model are required. As a result the histogram is formed as a sum of the proper decay, originating from the proper coincidence visible photons, and accidental decays, originating from the accidental coincidence photons. The corresponding formula for the probability distribution function for the time of fluorescent photon detection can be written as

\begin{equation}
P(t) = A \left(  p(t) + r \sum_{n=1}^{\infty} \left[ p(t + n t_0) + p(t - n \cdot t_0) \right] \right)  + b,
\label{eq:heralded_fluorescence_model}
\end{equation}
where $A$ and $b$ are respectively the experimental scaling factors related to the intensity and background, whereas $r$, reflecting the ratio of accidental to proper coincidences, is a factor correspoding to the probability with which the proper and accidental decays contribute to the process.

The observed range of data lies within $t\epsilon[-t_0, 3t_0]$, hence after substituting Eq. 1 into Eq. \ref{eq:heralded_fluorescence_model} the final model for the observed range is given by

\begin{multline}
P(t) = \frac{A}{\tau_N - \tau_R} 
\left(
r\left(\frac{e^{-\frac{t}{\tau_N}}}{e^{\frac{t_0}{\tau_N}}-1} 
- \frac{e^{-\frac{t}{\tau_R}}}{e^{\frac{t_0}{\tau_R}}-1}\right) \right. \\
+ \theta(t)\left(e^{-\frac{t}{\tau_N}} - e^{-\frac{t}{\tau_R}}\right) \\
+ \theta(t-t_0)r\left(e^{-\frac{t+t_0}{\tau_N}} - e^{-\frac{t+t_0}{\tau_R}}\right) \\
+ \left. \theta(t-2t_0)r\left(e^{-\frac{t+2t_0}{\tau_N}} - e^{-\frac{t+2t_0}{\tau_R}}\right) 
\right) + b.
\label{eq:model_single_conv_rep}
\end{multline}

\section{Signal to noise ratio (SNR)}
\label{appendix:SNR}

The heralded measurement scheme was applied. The heralded signal counts were estimated by multiplying the observed fluorescence signal counts, $n_f$, by the coincidence, $n_{coinc.}$, to visible, $n_{VIS}$, count rates ratio, giving the rate of heralded fluorescence counts per second
\begin{equation}
n_{sh} = \frac{n_{coinc.}}{n_{VIS}} \cdot n_{f}.
\label{eq:n_signal}
\end{equation}
The heralded dark counts were estimated as follows. The detected amount of dark counts (DC), $n_{DC}$, was multiplied by the probability of detecting a heralded DC,	$p_{herald.}$,

\begin{eqnarray}
n_{DCh} = p_{herald.} \cdot n_{DC},\\
\nonumber
p_{herald.} = \eta_{IR} \eta_{IR_{SSPD}}  \eta_{SPDC},
\end{eqnarray}
where $\eta_{IR}$ denotes the infrared photon fiber coupling efficiency, $\eta_{IR_{SSPD}}$ stands for the SSPD  quantum efficiency for the IR photon and $\eta_{SPDC}$ for the SPDC process efficiency per pulse. Hence, finally the SNR was calculated as a ratio

\begin{equation}
SNR = \frac{n_{sh} }{n_{DCh}}.
\end{equation}